\documentstyle[12pt,epsf]{article}
\def\beq{\begin{equation}} 
\def\eeq{\end{equation}} 
\def\bea{\begin{eqnarray}} 
\def\eea{\end{eqnarray}} 
\def\bq{\begin{quote}} 
\def\eq{\end{quote}}

\def\bq{\begin{quote}} 
\def\eq{\end{quote}}

\begin{document}

\baselineskip 15.5pt
\newcommand{\sheptitle}
{Yukawa Textures from Family Symmetry and Unification}

\newcommand{\shepauthor}
{B. C. Allanach$^1$, S. F. King$^2$, G. K. Leontaris$^{3}$ and S. Lola$^4$}

\newcommand{\shepaddress}
{$^1$Rutherford Appleton Laboratory, Chilton, Didcot, OX11 0QX, U.K.\\ 
$^2$Physics Department, University of Southampton\\Southampton, SO9 5NH, U.K.\\
$^3$Physics Department, University of Ioannina, PO Box 1186,
GR-45110 Ioannina, Greece \\
$^4$Theory Division, CERN, 1211 Geneva 23, Switzerland } 
\newcommand{\shepabstract}
{In this letter, we show how the combination
of both a gauged $U(1)_X$ family
symmetry and an extended vertical gauge symmetry 
in a single model, allows for the presence of
additional {\em Clebsch texture zeroes} \/in the fermion
mass matrices.
This, leads to new structures for the textures,
with increased predictivity, 
as compared to schemes with enhanced family symmetries only. 
We illustrate these ideas 
in the context of the Pati-Salam gauge group 
$SU(4)\otimes SU(2)_L\otimes SU(2)_R$ supplemented
by a $U(1)_X$ gauged family symmetry.  
In the case of symmetric mass matrices, two of the
solutions of Ramond, Roberts
and Ross that may not be obtained by family
symmetries only, are accurately reproduced. 
For non-symmetric textures, new structures arise
in models of this type. To distinguish between the solutions
in this latter case,
we performed a numerical fit to the charged fermion mass 
and mixing data. The best solution we found
allows
a fit with a total $\chi^2$ of 0.39, for three degrees of freedom.}

\begin{titlepage}
\begin{flushright}
CERN-TH/97-045 \\
IOA-TH 97-02 \\
RAL-TR-97-016 \\
SHEP 97-02 \\
{\tt hep-ph/9703361}
\end{flushright}
\vspace{.2in}
\begin{center}
{\large{\bf \sheptitle}}
\bigskip \\ \shepauthor \\ \mbox{} \\ {\it \shepaddress} \\ \vspace{.5in}
{\bf Abstract} \bigskip 
\end{center} 
\setcounter{page}{0}
\shepabstract
\end{titlepage}

In the recent years, there has been a lot of effort in trying to
understand  the pattern of quark and lepton masses and mixing angles.
In order to explain the observed hierarchies in the most
predictive way, it has been proposed
that zero textures in the Yukawa matrices
exist, like in the Fritzsch ansatz  \cite{Fritzsch}
and the Georgi-Jarlskog (GJ) texture  \cite{GJ}.
Ramond, Roberts and Ross (RRR)  \cite{RRR} have made a survey of possible
symmetric textures which are both consistent with data and involve
the maximum number of texture zeroes.  Such textures can arise due 
to a family symmetry group $G$ and some new heavy matter of mass $M$ which
transforms under $G$, resulting in effective non-renormalisable operators that
generate fermion masses  \cite{FN}. Such family symmetries arise in
most of the superstring models. A realisation of this picture has 
been provided by Ib\'a\~nez and Ross (IR) \cite{IR}, based on the MSSM extended
by a gauged family $U(1)_X$ symmetry with $\theta$ and $\bar{\theta}$ singlet 
fields with opposite $X$ charges, plus new heavy Higgs fields in vector 
representations. Anomaly cancellation  occurs via a Green-Schwarz-Witten
(GSW) mechanism, and the $U(1)_X$ symmetry is broken not far below the
string scale, generating Yukawa matrices of the form
\begin{equation}
\lambda^U = \left(\begin{array}{ccc}
\epsilon^8 & \epsilon^3 & \epsilon^4 \\
\epsilon^3 & \epsilon^2 & \epsilon \\
\epsilon^4 & \epsilon & 1 \\ \end{array}\right) 
,\ \ 
\lambda^D = \left(\begin{array}{ccc}
\bar{\epsilon}^8 & \bar{\epsilon}^3 & \bar{\epsilon}^4 \\
\bar{\epsilon}^3 & \bar{\epsilon}^2 & \bar{\epsilon} \\
\bar{\epsilon}^4 & \bar{\epsilon} & 1 \\ \end{array}\right) 
,\ \ 
\lambda^E = \left(\begin{array}{ccc}
\bar{\epsilon}^5 & \bar{\epsilon}^3 & 0 \\
\bar{\epsilon}^3 & \bar{\epsilon} & 0 \\
0  & 0  & 1 \\ \end{array}\right) 
\label{IR}
\end{equation}
These matrices resemble RRR solution 2  \cite{RRR} and
for $\epsilon \equiv  \bar{\epsilon}^2$, $\bar{\epsilon} = 
0.22$, they reproduce the known fermion mass hierarchy. However, the $23=32$
element in the down mass matrix tends to be large,
predicting a value for 
$|V_{cb}|$ that has to be lowered by the insertion
of coefficients or by a cancellation due 
to phases. On the other hand, one can
imagine  a different fit, that would naturally lead to a smaller $23$ element.
This can be achieved  by introducing a small parameter $\delta$ which
originates 
from some  flavour independent physics and appears as a factor in all
non-renormalisable
elements, so that e.g.\ the down quark mass matrix is modified to
\begin{equation}
\lambda^D = \left(\begin{array}{ccc}
\delta\bar{\epsilon}^8 & \delta\bar{\epsilon}^3 & \delta\bar{\epsilon}^4 \\
\delta\bar{\epsilon}^3 & \delta\bar{\epsilon}^2 & \delta\bar{\epsilon} \\
\delta\bar{\epsilon}^4 & \delta\bar{\epsilon} & 1 \\ \end{array}\right).
\label{deltaIR}
\label{deltaIRnumerical}
\end{equation} 
For $\bar{\epsilon}\approx 0.22$  and $\delta \approx 0.2$, in the down 
quark  mass matrix, one finds that all entries have naturally the correct
magnitude to reproduce the desired phenomenology.

However, in order to  naturally obtain such a contribution $\delta$ in
addition to the 
expansion parameter $\epsilon$ in the mass matrix entries, one has to go beyond
the MSSM\@. The basic new idea of our approach is to combine the ideas of 
gauged $U(1)_X$ family symmetry with unification in the form of an extended
vertical gauge symmetry.
As a concrete realisation of this idea we consider
a specific 
supersymmetric unified theory \cite{leo1} that can be derived from a superstring
model \cite{leo2}, based on the Pati-Salam
gauge group  $\mbox{SU(4)}\otimes \mbox{SU(2)}_L
\otimes \mbox{SU(2)}_R$ \cite{pati}. We emphasise that
the results can be generalised to
different groups; the important point is that the quarks and leptons
are unified into a common representation, leading to new Clebsch relations.
In this particular model, the 
left-handed quarks and leptons are accommodated in the
following representations (see \cite{leo1} for details)
\begin{equation}
{F^i}^{\alpha a}=(4,2,1)\,\,\,\,\, ,
{\bar{F}}_{x \alpha}^i = (\bar 4,1,2)
\end{equation}
where $\alpha=1,\ldots ,4$ is an SU(4) index, $a,x=1,2$ are
SU(2)$_{L,R}$ indices, and $i=1,2,3$ is a family index.  The Higgs
fields are contained in 
$h_{a}^x=(1,\bar{2},2)$
while the two heavy Higgs representations are
\begin{equation}
{H}^{\alpha b}=(4,1,2) ,
\;\; {\bar{H}}_{\alpha x}=(\bar{4},1,\bar{2})
\label{barH}
\end{equation}
The Higgs fields are assumed to develop VEVs $<H>=<\nu_H>\sim M_{GUT}, \ \
<\bar{H}>=<\bar{\nu}_H>\sim M_{GUT}$, leading to the symmetry breaking 
at $M_{GUT}$, $\mbox{SU(4)}\otimes \mbox{SU(2)}_L \otimes \mbox{SU(2)}_R$
$\longrightarrow$ 
$\mbox{SU(3)}_C \otimes \mbox{SU(2)}_L \otimes \mbox{U(1)}_Y$
in the usual notation. Under this 
symmetry breaking, the bidoublet Higgs field $h$ 
splits into two Higgs doublets 
$h_1$, $h_2$ whose neutral components subsequently
develop weak scale VEVs, $<h_1^0>=v_1$, $<h_2^0>=v_2 \label{vevs1}$ with $\tan
\beta \equiv v_2/v_1$. The relevant part of the superpotential involving matter
superfields is 
\begin{equation}
W =\lambda_1^{ij}\bar{F}_iF_jh + \lambda_2^{jk}\bar{F}_j{H}\theta_k
  +\mu hh +\cdots
\label{MSSMmatter}
\end{equation}
where $\theta_i$ are the superfields associated with the singlets. {}From 
Eq.\ref{MSSMmatter}  we find that the Yukawa couplings satisfy the boundary
conditions
\begin{equation}   
\lambda^{ij}_1 (M_{GUT}) \equiv \lambda^{ij}_U(M_{GUT}) 
= \lambda^{ij}_D (M_{GUT})=
\lambda^{ij}_E(M_{GUT}) = \lambda^{ij}_{\nu_D}(M_{GUT}),
 \label{boundary}
\end{equation}
When Eq.(\ref{boundary}) is applied to the third family ,
successful predictions for top, bottom and tau masses are obtained.
However these relations do not hold for the lighter families.
The standard way forward is to suppose that there is some family symmetry
which prohibits the renormalisable terms for the lighter families,
but permits them for the third family. The lighter fermion masses are
then accounted for by non-renormalisable operators whose order
is controlled by the family symmetry.

The non-renormalisable operators which were studied in
ref. \cite{422}, 
are formed from different group theoretical contractions of the 
indices  in
\begin{equation}
O^{\alpha \rho y w}_{\beta \gamma x z} \equiv F^{\alpha a}
\bar{F}_{\beta 
x} h^y_a \bar{H}_{\gamma z} H^{\rho w}.
\label{n1ops}
\end{equation}
and are of the form:
\begin{equation}
O_{ij}\sim (F_i\bar{F}_j
)h\left(\frac{H\bar{H}}{{M}^2}\right)^n+{\mbox h.c.} \label{op}
\end{equation}
where $M > M_{GUT}$ (in string models it can be associated with the string scale).
When $H, \bar{H}$ develop their VEVs, such operators will become effective Yukawa
couplings of the form $\, \bar{F}Fh$ with a small coefficient of order
$M_{GUT}^2/M^2$. These operators were previously written down independently 
of any particular family symmetry.
Now we extend this scheme by introducing a $U(1)_X$ family symmetry
which is broken  at a scale $M_X>M_{GUT}$ by the 
VEVs of the singlet fields $\theta$
and $\bar{\theta}$. 
To get an idea of the implications,  for simplicity 
 we assume that $(i)$ 
the singlet fields do not couple directly to the matter fields $F$ and $(ii)$
the the combination $H\bar{H}$ has a zero 
quantum number under the symmetry. The new operators are 
then modified from those in Eq.\ref{op}:
\begin{equation} \label{newop}
O_{ij}\sim (F_i\bar{F}_j)h\left(\frac{H\bar{H}}{M^2}\right)
\left(\frac{\theta^n \bar{\theta}^m}{{M'}^{n+m}}\right)+
{\mbox h.c.}
\label{newoperators} 
\end{equation}
where again ${M'}$ represents an energy scale which can be associated with
the string scale or the $U(1)_X$ breaking scale. The single power of $(H\bar{H})$
is present in every entry of the matrix and plays the role  of the factor 
$\delta$ in Eq.\ref{deltaIR}, while higher powers of $(H\bar{H})$ are suppressed.
In contrast to the MSSM case, here the factor $(H\bar{H})$ carries important
group theoretical Clebsch information. In fact Eq.\ref{newop} amounts to assuming
a sort of factorisation of the operators with the family hierarchies being
completely  controlled by the $\theta ,\bar{\theta}$ fields as in IR\@.
Here, $m$ and $n$
are dependent on $i,j$, while the splittings between different charge sectors
of the same family are controlled by the
Clebsch factors in $(H\bar{H})$. The Clebsch factors have of course a family
dependence, i.e.\ they depend on $i,j$.  
The relevant $n=1$ operators, appear in Table 1. 

We should comment on
the origin of the operators 
in Eq.\ref{newoperators}. There are two possible sources of these operators
which we can envisage: (1) the operators may emerge directly from
the superstring construction; (2) 
they may be generated from the 
effective field theory below the string scale. In case (1) the operators
will be suppressed by powers of the string scale, while in (2) the operators
will be suppressed by powers of the mass of some new 
heavy states in vector-like
representations which mix with the quark and lepton representations.
The first possibility will be addressed in \cite{NEW}.
Here, we would only like to remark that,
in this case, 
the larger contributions naturally arise 
by contracting fields belonging to the
same sector of the superstring (Neveu-Schwarz, or
Ramond sector), and that the operators
that arise in this way, lead to correct phenomenological
predictions.
As for the second possibility, this would require
heavy vector-like representations of matter fields 
$\psi^{(x)} +\bar{\psi}^{(-x)}$ 
where the $X$ charges are shown in parentheses.
Flavour mixing occurs via insertions of $\theta$ and $\bar{\theta}$
fields leading to the so called spaghetti diagrams recently discussed
in the context of the MSSM \cite{spaghetti}.
In the case of the Pati-Salam model, additional heavy Higgs
fields in adjoint representations may generate
the operators in Eq.\ref{newoperators}.
An example of a spaghetti diagram which can yield a 23 
entry in the Yukawa matrix is shown in Fig.1. 
Note that the $H\bar{H}$ pair will be in the
Pati-Salam representation of the heavy $\Sigma$ field which does
not change flavour since it has zero $X$ charge (in parentheses).
Flavour changing occurs only in the heavy $\psi,\bar{\psi}$ sector
(where these fields have the same Pati-Salam representations
as the chiral matter fields $F_i,\bar{F}_j$)
via insertions of the $\bar{\theta}$ fields.

\begin{figure}
\begin{center}
\leavevmode   
\hbox{\epsfxsize=3in
\epsfysize=2in
\epsffile{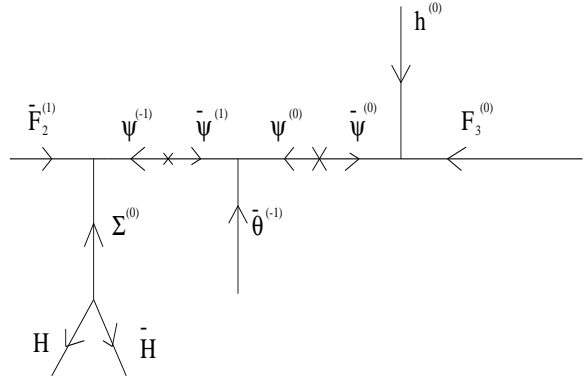}}
\end{center}
\caption{A Pati-Salam spaghetti diagram.}
\label{Fig.1}
\end{figure}

\begin{table} 
\begin{center}
\begin{tabular}{|c|c|c|c|c|} \hline
 & $Q \bar{U} h_2$ & $Q \bar{D} h_1$ & $L \bar{E} h_1$ & $L \bar{N}
h_2$
\\ \hline
$O^A$ &1 & 1 & 1 &1 \\ 
$O^B$ &1 & -1& -1 &1 \\ 
$O^C$ &$\frac{1}{\sqrt{5}}$ & $\frac{1}{\sqrt{5}}$ 
& $\frac{-3}{\sqrt{5}}$ &$\frac{-3}{\sqrt{5}}$  \\ 
$O^D$ &$\frac{1}{\sqrt{5}}$ & $\frac{-1}{\sqrt{5}}$
& $\frac{3}{\sqrt{5}}$ & $\frac{-3}{\sqrt{5}}$\\ 
$O^G$ & 0 & $\frac{2}{\sqrt{5}}$ & $\frac{4}{\sqrt{5}}$ & 0 \\
 $O^H$ & 4/5 & 2/5 & 4/5 & 8/5 \\
$O^K$ & 8/5 & 0 & 0 & 6/5 \\
$O^M$ &0 & $\sqrt{2}$ & $\sqrt{2}$ &0 \\ 
$O^N$ &2&0&0&0\\ 
$O^R$ &0&$\frac{8}{5}$& $\frac{6}{5}$& 0 \\
$O^W$ &0 & $\sqrt{\frac{2}{5}}$& -3$\sqrt{\frac{2}{5}}$&0\\
$O^S$ & $\frac{8}{5 \sqrt{5}}$ & $\frac{16}{5 \sqrt{5}}$ &
$\frac{12}{5 \sqrt{5}}$ & $\frac{6}{5 \sqrt{5}}$ \\
\hline
\end{tabular}
\end{center}
\label{tab:subset}
\caption{{\small The
$n=1$ operators with
Clebsch coefficients as shown.}}
\end{table}

Note that Table 1 includes cases of {\em zero Clebsch coefficients}, 
where the  contribution to the up-type matrix, for example, is
precisely zero. Similarly there are zero Clebsch coefficients for the
down-type
quarks (and charged leptons).  The existence of such zero Clebsch coefficients
enables us to obtain textures with different places for zeroes in the up and
down mass matrices, something that is not easy in the case that only a flavour
symmetry is included in the model. Apart from the zero Clebsch coefficients
another advantage of combining quark-lepton
unification with a family symmetry is that one can also account for the mass
splitting within a particular family, since now the {\em same expansion
 parameter}
\/appears in both the up and down quark mass matrices. 

To see how all this occurs,
let us  take a specific example. 
Consider the symmetric $n=1$ operator texture,
\begin{equation}
\lambda = \left(\begin{array}{ccc}
0 & O^M & O^{N} \\
O^M & O^W + s.d. & O^{N} \\
O^{N}  & O^{N}  & O_{33} \\ \end{array}\right) 
\label{RRR5operators}
\end{equation}
where $O_{33}$ is the renormalisable operator
and ${s.d.}$ stands for a
sub-dominant 
operator with a suppression factor compared to the other dominant operator in
the 
same entry. Substituting the Clebsch coefficients from Table 1 we arrive at the
following
$\lambda^{U,D,E}$  Yukawa matrices, at the GUT scale,
\begin{equation}
\lambda^U = \left(\begin{array}{ccc}
0 & 0  & 2\lambda_{13}^U \\
0 & \lambda_{22}^U & 2\lambda_{23}^U \\
2\lambda_{13}^U  & 2\lambda_{23}^U   & 1 \\ \end{array}\right),\; 
\lambda^D = \left(\begin{array}{ccc}
0 & \sqrt{2}\lambda_{12}^D & 0\\
\sqrt{2}\lambda_{12}^D &\sqrt{\frac 25} \lambda_{22}^D & 0 \\
0  & 0    & 1 \\ \end{array}\right), \; 
\lambda^E = \left(\begin{array}{ccc}
0 & \sqrt{2}\lambda_{12}^D & 0\\
\sqrt{2}\lambda_{12}^D & 3 \sqrt{\frac 25}\lambda_{22}^D & 0 \\
0  & 0    & 1 \\ \end{array}\right) 
\label{RRR5leptoncomponents}
\end{equation}
where $\lambda_{22}^D$ and $\lambda_{22}^E$ arise from the dominant $O^W_{22}$
operator and $\lambda_{22}^U$ comes from a 
sub-dominant operator that is relevant
because of the texture zero Clebsch 
in the up sector of $O^W_{22}$. The zeroes in
the matrices correspond to those of the RRR solution 5, 
but of course in our case
they arise from the Clebsch zeroes 
rather than from a family symmetry. The numerical values 
corresponding to RRR solution 5 are,
\begin{equation}
\lambda^U = \left(\begin{array}{ccc}
0 & 0 & 2\times 10^{-3} \\
0 & 3\times 10^{-3} & 3\times 10^{-2} \\
2\times 10^{-3} & 3\times 10^{-2}  & 1 \\ \end{array}\right) 
,\ 
\lambda^D = \left(\begin{array}{ccc}
0 & 5\times 10^{-3}& 0 \\
5\times 10^{-3} & 2\times 10^{-2} & 0 \\
0 & 0 & 1 \\ \end{array}\right) 
\label{RRR5numerical}
\end{equation}
In our model the hierarchy $\lambda_{22}^U << \lambda_{22}^D$ is explained by
a Clebsch zero and a suppression factor of the sub-dominant operator.
Using Eq.\ref{RRR5numerical} we can read off the values of the couplings which 
roughly correspond to a unified matrix of effective (and dominant) couplings:
\begin{equation}
\lambda_{ij} = \left(\begin{array}{ccc}
0 & 3\times 10^{-3}& 1\times 10^{-3} \\
3\times 10^{-3} & 2\times 10^{-2} & 2\times 10^{-2} \\
1\times 10^{-3} & 2\times 10^{-2} & 1 \\ \end{array}\right) 
\label{unifiednumerical}
\end{equation}
where we have extracted the Clebsch factors. 
Thus, the Clebsch factors reproduce the values in Eq.\ref{RRR5numerical}
required by phenomenology for the dominant effective couplings displayed in
Eq.\ref{unifiednumerical}.

The case we are examining is different from the IR analysis in two aspects:   
$(a)$ the fermion mass 
matrices of different charge sectors have the same origin, and thus the same
expansion
parameter and $(b)$ all variations between these sectors arise 
from Clebsch factors. The structure of the mass matrices is again determined 
by a family symmetry 
$U(1)_{X}$, but now 
the charge assignment of the various 
states are as in Table~\ref{table:1}.
\begin{table}[h]
\centering
\begin{tabular}{|c |cccccccccc|}\hline
   &$ Q_i$ & $u^c_i$ &$ d^c_i$ &$ L_i$ & $e^c_i$ & $\nu^c_i$ &
$h_1$ &
$ h_2$ & $H$ & $\bar{H}$   \\
\hline
  $U(1)_{X}$ & $\alpha_i$ & $\alpha _i$ & $\alpha _i$  & $\alpha_i$
& $\alpha_i$ & $\alpha_i $ & $-\alpha_3-\alpha _3$ &  $-\alpha_3-\alpha _3$
& x & -x 
\\
\hline
\end{tabular}
\caption{{\small $U(1)_{X}$ charges for symmetric
textures. }
\label{table:1}}
\end{table}

The need to preserve $SU(2)_L$
invariance requires left-handed up and down quarks (leptons)
to have the same charge. This, plus the additional
requirement of symmetric
matrices, indicates that all quarks (leptons) of the same $i$-th
generation transform with the same charge $\alpha _i$.  
Finally, lepton-quark unification under 
$SU(4) \otimes SU(2)_L \otimes SU(2)_R$  indicates that
quarks and leptons of the same family have the same charge.
The full anomaly free Abelian group involves an additional family
independent component, $U(1)_{FI}$, and with this freedom
$U(1)_{X}$ is made traceless without any loss of 
generality, giving $\alpha_1=-(\alpha_2+\alpha_3)$.

If the light Higgs $h_{2}$, $h_{1}$, responsible for the up and
down quark masses respectively, arise from the same bidoublet
$h=(1,2,2)$, then they have the same $U(1)_X$ charge so that only 
the 33 renormalisable Yukawa coupling to $h_{2}$, $h_{1}$ is
allowed, and only the 33 element of the associated mass matrix
will be non-zero.  The remaining entries are generated when the
$U(1)_X$ symmetry is broken,
via Standard Model singlet fields, which can be either chiral
or vector ones.

In our case,  all charge and mass
matrices have identical structure under the
$U(1)_{X}$ symmetry, since all known fermions are
accommodated in the same multiplets of the gauge group.
The charge matrix is of the form
\begin{eqnarray}
\left(
\begin{array}{ccc}
-2 \alpha_2 - 4 \alpha_3 & 
-3 \alpha_3 & -\alpha_2 - 2\alpha_3 \\
-3 \alpha_3 & 2(\alpha_2-\alpha_3) 
& \alpha_2 - \alpha_3 \\
-\alpha_2 -2 \alpha_3 & \alpha_2 - \alpha_3 & 0
\end{array}
\right)
\end{eqnarray}
Then, including the
common factor $\delta$ which multiplies
all non-renormalisable entries,
the following hierarchy of dominant effective Yukawa couplings is obtained
\begin{eqnarray}
\lambda\approx 
\left(
\begin{array}{ccc}
\delta \epsilon^{\mid 2+6a \mid } &
\delta \epsilon^{\mid 3a \mid } &
\delta \epsilon^{\mid 1+3a\mid }
\\
\delta \epsilon^{\mid 3a \mid } &
\delta \epsilon^{ 2 } &
\delta \epsilon \\
\delta \epsilon^{\mid 1+3a \mid } &
\delta \epsilon & 1
\end{array}
\right)
, \; \; \; \;
\label{eq:massu}
\end{eqnarray}
where we have used a vector-like pair of $SU(4) \otimes SU(2)_L \otimes
SU(2)_R$ singlets to Higgs the family symmetry. 
Here
$\epsilon=(<\theta>/M)^{\mid\alpha_2-\alpha_3\mid}$
where $M$ is the unification mass
scale  which governs the higher dimension operators,
while a unique charge combination $a=\alpha_3/(\alpha_2-\alpha_3)$
appears in the exponents of all matrices, as a result
of quark-lepton unification. 
In Eq.\ref{eq:massu} the
flavour independent suppression factor $\delta=\langle H \bar{H}\rangle/M^2$
results from the
operators in Eq.\ref{RRR5operators}.
The factor of $\delta $ actually helps the numerical fit and
setting $a=1$, $\epsilon = 0.22$ and  $\delta \approx 0.2$ successfully 
reproduces the desired values of the elements of the Yukawa matrix
in Eq.\ref{unifiednumerical}. Thus, starting with all of the
``fundamental'' Yukawa couplings\footnote{i.e.\ dimensionless couplings in the
full 
SU(4)$\otimes$SU(2)$_L\otimes$SU(2)$_R\otimes$U(1)$_X$ gauge theory.} $\sim
1$, we can reproduce the correct phenomenology of charged fermion masses and
mixings.

Let us now look at the case of non-symmetric textures\footnote{Asymmetric
textures have been discussed in a different
framework in \cite{blr}.},
with an additional zero in the 23 entry.
It turns out \cite{422,NEW} that it is optimal to
consider a scheme in which
the dominant operators in the Yukawa matrix are $O_{33}$, $O_{32}^C$,
$O_{22}^W$, $O_{21}, \tilde{O}_{21}$ and $O_{12}$, where the last 
three operators are left general and
will be specified later.
We are also aware from
the analysis in ref. \cite{422} that $O_{12}$ must have a zero Clebsch
coefficient in the up sector. A combination of two operators must
then provide a non-zero $O_{21}$ entry to generate a big enough
$V_{ub}$, while 
an additional much more suppressed operator elsewhere in
the Yukawa matrix gives the up quark a small mass.
At $M_{GUT}$, the Yukawa matrices that we can form
from the operators listed above, are of the form
\begin{equation}
\lambda^I = \left[ \begin{array}{ccc} 0 & H_{12} e^{i \phi_{12}}
x_{12}^I 
& 0
\\ H_{21} x_{21}^I e^{i \phi_{21}}+ \tilde{H}_{21} \tilde{x}_{21}^I
e^{i\tilde{\phi}_{21}} & H_{22} x_{22}^I e^{i \phi_{22}} &
0 \\
0 & H_{32} x_{32}^I e^{i \phi_{32}} & H_{33} e^{i \phi_{33}} \\
\end{array}\right], \label{dom}
\end{equation}
where only the dominant operators are listed
and complex phases have been taken into
account. The $I$ superscript
labels the charge sector and $x_{ij}^I$ refers to the Clebsch
coefficient relevant to the charge sector $I$ in the $ij^{th}$
position. $\phi_{ij}$ are unknown phases and $H_{ij}$ is the magnitude
of the effective dimensionless Yukawa coupling in the $ij^{th}$
position. Any subdominant operators that we introduce will be denoted
below by a prime and it should be borne in mind that these will only
affect the up matrix. So far, the known Clebsch coefficients are
\begin{eqnarray}
x_{12}^U = 0  & & \nonumber \\
x_{22}^U = 0 & x_{22}^D = 1 & x_{22}^E = -3 \nonumber \\
x_{32}^U = 1 & x_{32}^D = -1 & x_{32}^E = -3.
\label{cleb}
\end{eqnarray}
We have just enough freedom in rotating the phases of $F_{1,2,3}$ and
$\bar{F}_{1,2,3}$ to get rid of all but one of the phases in
Eq.\ref{dom}.
When the subdominant operator is added, the Yukawa matrices are
\begin{eqnarray}
\lambda^U &=& \left[\begin{array}{ccc} 0 & 0 & 0\\
H_{21}^U e^{i \phi_{21}^U} & {H_{22}}' e^{i {\phi_{22}}'} & 0 \\
0 & H_{32} x_{32}^U & H_{33} \\ \end{array}\right] \nonumber \\
\lambda^D &=& \left[\begin{array}{ccc} 0 & H_{12} x_{12}^D & 0\\
H_{21}^D & H_{22} x_{22}^D & 0 \\
0 & H_{32} x_{32}^D & H_{33} \\ \end{array}\right] \nonumber \\
\lambda^E &=& \left[\begin{array}{ccc} 0 & H_{12} x_{12}^E & 0\\
H_{21}^E & H_{22} x_{22}^E & 0 \\
0 & H_{32} x_{32}^E & H_{33} \\ \end{array}\right], \label{mxyuks}
\end{eqnarray}
where we have defined
\begin{eqnarray}
H_{21}^U e^{i \phi_{21}^U} &\equiv& H_{21}
x_{21}^U e^{i\phi_{21}}
+ \tilde{H}_{21} \tilde{x}_{21}^U e^{i\tilde{\phi}_{21}}  \nonumber \\
H_{21}^{D,E} &\equiv& H_{21}
x_{21}^{D,E} e^{i\phi_{21}}
+ \tilde{H}_{21} \tilde{x}_{21}^{D,E} e^{i\tilde{\phi}_{21}} 
\end{eqnarray}
We may now remove ${\phi_{22}}'$ by phase transformations upon
$\bar{F}_{1,2,3}$ but $\phi_{21}^U$ may only be removed by a phase
redefinition of $F_{1,2,3}$, which would alter the prediction of the
CKM matrix $V_{CKM}$. Thus, $\phi_{21}^U$ is a physical phase, that is
it cannot be completely removed by phase rotations upon the fields.
Once the operators $O_{21},\tilde{O}_{21}, 
O_{12}$ have been chosen, the Yukawa
matrices at  $M_{GUT}$ including the phase in the CKM matrix are
therefore
identified with $H_{ij}, {H'}_{22},\phi_{21}^U$.

This type of non-symmetric texture can be described
by a structure of the kind,
\begin{equation}
\lambda = \left(\begin{array}{ccc}
\delta {\epsilon}^{big}& \delta {\epsilon}^3 & \delta {\epsilon}^{big} \\
\delta {\epsilon}^3 & \delta {\epsilon}^{1or2} & \delta {\epsilon}^{big} \\
\delta {\epsilon}^{big} & \delta {\epsilon} & 1 \\ \end{array}\right) 
\label{deltanonsym}
\end{equation}
where we identify $\epsilon \equiv \lambda = 0.22$
and set $\delta \approx 0.1$.

Let us numerically analyse the non-symmetric
case we have been discussing, since it goes beyond
solutions that have been discussed so far in the literature.
This is done by performing a global fit of possible 
textures, arising from different
operator assignments, to
$m_e,m_\mu$, 
$m_u$,
$m_c$,
$m_t$,
$m_d$,
$m_s$,
$m_b$,
$\alpha_S(M_Z)$,
$|V_{ub}|$, $|V_{cb}|$ and
$|V_{us}|$ using $m_\tau$ as a constraint.
The values of the 8 parameters
introduced in Eq.\ref{mxyuks}
($\phi_{21}^U \equiv \phi$, $H^U_{21}\equiv {H_{21}}'$, 
$H^D_{21}\equiv H_{21}$, 
${H_{22}}', H_{22}$, $H_{12}$,
$H_{32}, H_{33}$),
plus $\alpha_S$ at the GUT scale are determined by the fit.

The matrices $\lambda^I$ are diagonalised numerically and 
$|V_{ub}(M_{GUT})|, |V_{us}(M_{GUT})|$ are determined by 
$V_{CKM} = {V_U}_L {V_D}_L^\dagger$,
where ${V_U}_L, {V_D}_L$ are the matrices that act upon the
$(u,c,t)_L$ and $(d,s,b)_L$ column
vectors respectively to transform from the weak eigenstates to the mass
eigenstates of the quarks. We use the boundary conditions
$\alpha_1(M_{GUT})=\alpha_2(M_{GUT})=0.708$, while
$\lambda_{u,c,t,d,s,b,e,\mu,\tau}$, $|V_{us}|$ and $|V_{ub}|$ are
 run\footnote{All
renormalisation running in this paper is one loop and in the
$\overline{\mbox{MS}}$ scheme. The relevant renormalisation group
equations (RGEs)
are listed in ref. \cite{422}.} from $M_{GUT}$ to 170 GeV$\approx m_t$
using the
RGEs for the MSSM\@. 
The $\lambda_i$ are  evolved to their
empirically derived running masses using 3 loop QCD$\otimes$1 loop
QED  \cite{422}. 
$m_\tau^{e}$ and $\lambda_\tau^{p}(m_\tau)$ then\footnote{The
superscript $e$ upon masses, mixing angles or diagonal Yukawa
couplings denotes an empirically derived value, whereas the
superscript $p$ denotes the prediction of the model for the 
particular fit
parameters being tested.}
fix $\tan \beta$ through
the relation~\cite{Andersonetal}
$\cos \beta = \frac{\sqrt{2} m_{\tau}^e (m_\tau)}{v \lambda_\tau^p
(\lambda_\tau)}$,
where $v=246.22$ GeV is the VEV of the Standard Model Higgs.
There are twelve data points and nine parameters so we
have three degrees of freedom (dof). The parameters are all varied until the
global $\chi^2 / \mbox{dof}$ is minimised. The data used (with
1$\sigma$ errors quoted) is taken from \cite{databook}.
For
\begin{equation}
O_{12} \equiv O^R, \; \; \;
O_{21}+\tilde{O}_{21} \equiv O^M+O^A
\end{equation}
we obtain the best fit. The result appears in
Table~\ref{tab:res1}.

\begin{table}
\begin{center}
\begin{tabular}{|c|c|c|c|c|c|c|} 
\hline
 $H_{22}/10^{-2}$ & 
 $H_{12}/10^{-3}$ & 
 $H_{21}/10^{-3}$ & 
 $\cos \phi$  & 
 $H_{33}$ & 
 ${H_{22}}'/10^{-3}$  & 
 ${H_{21}}'/10^{-3}$  \\
\hline
2.69 & 2.13 & 1.76 & 0.20 & 1.05 & 1.87 & 1.63 \\
\hline
\end{tabular}

\vspace*{0.3 cm}
\begin{tabular}{|c|c|c|c|c|}
\hline
 $\alpha_S(M_Z)$ &  
$m_d$/MeV & 
$m_s$/MeV & 
$m_c$/GeV & 
$m_b$/GeV \\
\hline
0.118 & 8.07 & 154 & 1.30 & 4.25 \\
\hline
\hline
$m_t^{phys}$/GeV &
$|V_{us}|$ & 
$|V_{ub}|/10^{-3}$ &
$\tan \beta$ 
 & {\mbox{\boldmath{$\chi^2$}}{\bf /dof}}  \\
\hline
 180 & 0.2215 &
3.50 & 58.3 & {\bf 0.13} \\
\hline
\end{tabular}
\end{center}
\caption{{\small Results of best-fit analysis on models with $n=1$ operators.}}
\label{tab:res1}
\end{table}

Out of 16 possible models that fit the texture required by
Eqs.\ref{cleb},\ref{dom}, 11 models fit the data with
$\chi^2$/dof$<3$, 5 with
$\chi^2$/dof$<2$ and 3 with 
$\chi^2$/dof$<1$.
The operators listed as $O_{12}$,$O_{21}$,$\tilde{O}_{21}$ describe the
structure of the
models and the entries $H_{22}$, $H_{12}$, $H_{21}$, $\cos \phi$, 
$H_{33}$,
${H_{22}}'$, ${H_{21}}'$ are the GUT scale input parameters of the best
fit values of the model. The estimated 1$\sigma$ deviation in
$\alpha_S(M_Z)$ from the fits is $\pm 0.003$ and the other parameters
are constrained to better than 1\% apart from $\cos \phi$, whose
1$\sigma$ fit 
errors often cover the whole possible range.
We conclude that the $\chi^2$ test has some discriminatory power,
since if all of the models were equally good, we would
statistically expect to have 11 models with $\chi^2$/dof$<1$, 3 models
with $\chi^2$/dof$=1-2$ and 2 models with $\chi^2=2-3$ out of the 16
tested.

Summarising, we have combined the idea of a gauged $U(1)_X$ family 
symmetry with that of quark-lepton unification within the framework 
of a string-inspired Pati-Salam model. The non-renormalisable operators 
are composed of a factor $(H\bar{H})$ and a factor involving the singlet 
fields $\theta,\bar\theta$ as in Eq.\ref{newoperators}.
The singlet fields $\theta,\bar\theta$ break the $U(1)_X$
symmetry and provide the horizontal family hierarchies while the $H,\bar H$
fields break the SU(4)$\otimes$SU(2)$_L\otimes$SU(2)$_R$ symmetry and give the
vertical  splittings arising from group theoretic Clebsch relations between
different charge sectors.  We have studied both symmetric and
asymmetric mass textures.
In the first case,
as an example, we have shown how one of the RRR textures can be
reproduced by the structure in Eq.\ref{eq:massu}, where
$\delta$ results from the factor $\langle H \bar{H}\rangle/M^2$ 
and $\epsilon$ results from
the factor $(<\theta>/M)^{\mid\alpha_2-\alpha_3\mid}$ in the operators of
Eq.\ref{RRR5operators}.
This results to an accurate reproduction of 
RRR solution 5 which cannot be reproduced by family symmetry alone.
The Giudice ansatz~\cite{Giudice} (RRR solution 3), which also cannot normally
be achieved, 
can similarly 
be reproduced by our scheme. 
Our approach has also been extended to non-symmetric textures,
which are motivated by particular superstring constructions.
For these, we performed a numerical analysis since 
they are beyond the symmetric RRR regime
and found that from the possible textures, one of them allows
a low energy fit to the data, with a total $\chi^2$ of 0.39
for three degrees of freedom.
All the textures that we present, rely on
the existence of Clebsch zeroes, which are a feature of the operators
in our model. Moreover, the combination of 
family symmetry and unification
has the additional advantage that quark and lepton masses are related by
Clebsch factors, as in the Georgi-Jarlskog scheme,
which improves the overall predictive power of the approach. 
More detailed issues, including the
 superstring construction itself, are deferred
to a longer publication  \cite{NEW}.

\pagebreak

\begin{center}
{\bf Acknowledgments}
\end{center}
B.C.A. would like to thank J.Holt for advice on the $\chi^2$ test.
The work of S.F.K. is partially supported by PPARC grant number GR/K55738.
S.L. would like to acknowledge the Theory Group at the University of 
Southampton for an one-month PPARC funded Research Associateship
which greatly facilitated this research.
The work of S.L. is funded by a Marie Curie Fellowship
(TMR ERBFMBICT-950565) and at the initial stages by
$\Pi$ENE$\Delta$-91$\Delta$300. 
The work of G.K.L. is
partially supported by $\Pi$ENE$\Delta$-91$\Delta$300.


\end{document}